\pdfoutput=1
\documentclass[pra,aps,twocolumn]{revtex4}
\usepackage{graphicx}
\usepackage{amsmath}
\usepackage{color}
\usepackage{soul}

\begin{document}

\title{Heat-induced soliton self-frequency redshift in the ultrafast nonlinear dynamics of active plasmonic waveguides}
\author{Ambaresh Sahoo$^1$, Andrea Marini$^2$ and Samudra Roy$^{1,*}$}
\affiliation{$^1$Department of Physics, Indian Institute of Technology Kharagpur, West Bengal 721302, India}
\email{samudra.roy@phy.iitkgp.ernet.in}
\affiliation{$^2$Department of Physical and Chemical Sciences, University of L'Aquila, Via Vetoio, 67100 L'Aquila, Italy}

\begin{abstract}
We investigate the ultrafast nonlinear dynamics of light emitted in an active plasmonic waveguide composed of a thin film of gold sandwiched by two silicon layers immersed in externally pumped Al$_2$O$_3$:Er$^{3+}$. We model optical propagation in such a dissipative system through a generalized cubic Ginzburg-Landau equation accounting for the amplification of the active medium and the effect of absorption and thermo-modulational nonlinearity of gold. We find that heating heavily affects the propagation of temporal dissipative solitons in such a plasmonic waveguide by producing a soliton self-frequency redshift accompanied by soliton deceleration in the time domain. By adopting a semi-analytical variational approach, we evaluate the dependence of the self-induced redshift by deriving a set of coupled differential equations for the pulse parameters. These equations provide physical insight into the complex nonlinear dynamics through simple approximate analytic expressions for temporal and frequency shifts. Such analytical predictions are found in excellent agreement with direct numerical simulations of the generalized cubic Ginzburg-Landau equation. Our results provide a general understanding of ultrafast nonlinear dynamics in gold-based active plasmonic waveguides, as in particular the spectral shaping properties of propagating optical pulses.   
\end{abstract}

\maketitle

\section{Introduction}

Plasmonic materials exploit the tight confinement and large field enhancement produced by plasmons to confine light down to the nanoscale. Surface plasmon polaritons (SPPs) arise from the coupling of
photons and plasma waves excited at metal-dielectric interfaces \cite{Ritchie1968,Zayats2005}, enabling the enhancement of light-matter interaction and optical nonlinearity \cite{Kauranen2012}, nano-control \cite{Barnes2003}, and on-chip photonic devices \cite{Ozbay2006,Schmidt}. Furthermore, SPPs are particularly important also in medical applications \cite{GobinNanoLett2007}, biosensing \cite{AnkerNatMat2008}, and imaging \cite{KawataNatPhot2009}. Owing to the high confinement provided by SPPs, a plethora of nonlinear effects arising from metals are enhanced, e.g. nonlocal ponderomotive forces  \cite{DavoyanOL2011}, interband thermo-modulation \cite{Conforti,MariniNewJPhys2013}, second and third harmonic generation \cite{GinzburgNewJPhys2013}. The enhanced Kerr nonlinearity can also lead to plasmon-soliton formation \cite{Feigenbaum2007} and nanofocusing \cite{DavoyanPRL2010}. Such nonlinear applications are however hampered by the inherently large ohmic losses of metals \cite{Khurgin2015}. A way to tackle losses consists of exploiting self-induced transparency plasmon solitons to suppress interband absorption in gold-based plasmonic nanowires \cite{Marini2}. Furthermore, embedding amplifying media in plasmonic devices can compensate for ohmic losses and lengthen the propagation of SPPs \cite{GatherNature2010}. 

Nonlinear dynamics in amplifying plasmonic waveguides can be further utilized to obtain transverse localization of dissipative plasmon solitons (DPSs) \cite{MariniPRA2010,MariniOE2010,Ferrando2017}. Temporal localization can also be achieved by DPSs in active plasmonic waveguides through the double balance between dispersion and nonlinearity and between gain and losses~\cite{Akhmediev}. Hybrid gold-silicon plasmonic waveguides surrounded by an active gain medium can support DPS formation. However, the nonlinear temporal dynamics in gold-based plasmonic devices is heavily influenced by light-induced heating of the conduction electrons \cite{rosei,guerrisi}. In practice, the gold nonlinear susceptibility at optical and near-infrared frequencies is determined by the interband transitions around the X, L points in the first Brillouin zone and thermo-modulational effects arise from Fermi smearing of the electronic energy distribution in the conduction band. Theoretical and experimental investigations on the temporal dynamics of such a process indicate that the nonlinear response of gold is characterised by a delayed mechanism resulting from the different time scales of electron-electron and electron-phonon scattering \cite{rosei,guerrisi,bache}. Such a delayed response is responsible for a strong deceleration of femtosecond optical pulses accompanied by redshift in the frequency domain \cite{MariniNewJPhys2013}. Although nonlinear dynamics of amplified SPPs has already been investigated extensively over recent years \cite{Berini,Smirnova}, the effect of gold thermo-modulational nonlinearity on SPP lasing remains unexplored. 

Here we model the dynamics of amplified SPPs by a complex Ginzburg-Landau equation (CGLE) to describe pulse propagation in an active plasmonic waveguide composed of a thin film of gold sandwiched by two silicon layers immersed in externally pumped Al$_2$O$_3$:Er$^{3+}$. We further develop a semi-analytical variational technique to describe the physics underpinning the excitation of DPSs \cite{Bondeson,Kaup,Cerda}, thus enabling us to predict the evolution of DPS parameters. Thus, we describe the evolution dynamics of various pulse parameters predicting accurately the DPS spectral red-shift ignited by the thermo-modulational nonlinearity of gold along with the accompanied deceleration. Finally we compare our analytical predictions with direct numerical solutions of the CGLE, demonstrating that the variational approach provides precise qualitative and quantitative hints on DPS propagation. 

\begin{figure}[t]
\centering
\begin{center}
\includegraphics[width=0.5\textwidth]{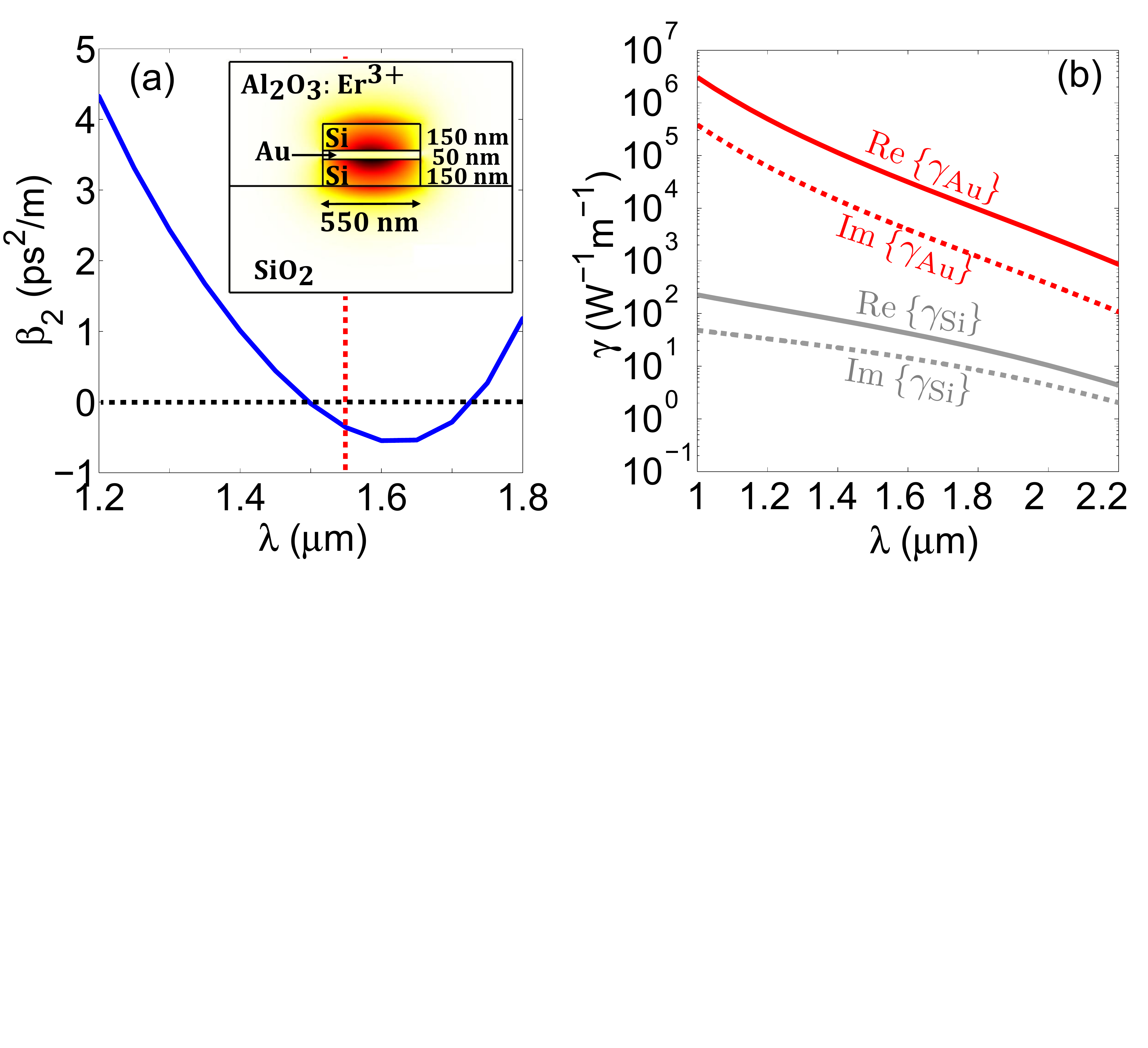}
\caption{(Color online) A hybrid gold-silicon active waveguide composed of a gold nanostripe sandwiched by silicon and surrounded by an active gain medium. The field distribution of the fundamental TM plasmonic mode at $\lambda_0=1.55 \ \mu$m is depicted in (a). The solid blue line represents the group velocity dispersion (GVD) of the fundamental TM-mode under consideration \cite{COMSOL}. (b) The complex third order nonlinear coefficients of silicon ($\gamma_{\rm Si}$) and gold ($\gamma_{\rm Au}$) corresponding to the mode in (a) are plotted in semi-logarithmic scale (real and imaginary parts are plotted in full and dashed curves, respectively).}  
\label{DS_model}
\end{center}
\end{figure}

\section{Setup}

Owing to the strong absorption of gold, the propagation of pulses in passive systems (such as a plasmonic structure containing a gold nanowire surrounded by silica glass) can not be sustained over a propagation distance longer than few microns \cite{MariniNewJPhys2013}. To overcome this limitation and obtain SPP pulse propagation over longer distances it is required to embed an externally pumped amplifying medium in the system feeding continuously energy and counterbalancing absorption, which is the underlying principle of active plasmonic waveguides. In order to investigate SPP dynamics in active dissipative systems, we consider a hybrid gold-silicon waveguide structure composed of a gold nanostripe of cross section of $50$nm$\times 550$nm depicted in the inset of Fig.~\ref{DS_model}(a) along with the field distribution of the fundamental TM plasmonic mode at wavelength $\lambda=1.5~\mu$m. The mode field is tightly bound to the metal surface and is semi-circular in shape. The GVD profile of the proposed structure is depicted in Fig.~\ref{DS_model}(a) by the solid blue line for the wavelength range $0.95\ \mu$m to $2.25\ \mu$m. In Fig.~\ref{DS_model}(b) we plot (in the semi-log scale) the complex third-order nonlinear coefficients of silicon ($\gamma_{\rm Si}$) and gold ($\gamma_{\rm Au}$). The calculation of such nonlinear coefficients can be done either by developing a perturbation theory for the nonlinear terms accounting for the surface nonlinearity or through the reciprocity theorem \cite{MariniPRA2011,Afshar}, both approaches leading to the expressions reported in Eqs. (\ref{NLCoeffEqs}). We emphasize that $|\gamma_{\rm Si}|<<|\gamma_{\rm Au}|$ in the spectral range considered in Fig.~\ref{DS_model} (b), and thus neither Kerr nor Raman effects of silicon play any significant role. Hence, for the sake of simplicity, in our calculations below we disregard the silicon nonlinearity.

\section{Thermo-modulational nonlinearity of gold}

The ultrafast temporal dynamics of photo-generated electrons is characterized by different time scales dictated by electron-electron and electron-phonon scattering \cite{MariniNewJPhys2013}. Indeed, while electron-electron scattering happens at the femtosecond time scale leading to a thermalization time $\tau_{\rm th}\simeq 300$ fs, electron-phonon thermalization responsible for relaxation occurs over a time scale of the order of $\tau_{\rm r}\simeq 1$ ps. Thus, the electron temperature can become much larger than the lattice temperature over the $100$ fs scale while relaxing to the lattice temperature over the ps scale. Such an ultrafast dynamics is phenomenologically captured by the two-temperature model \cite{MariniNewJPhys2013}, which accounts for the transient variation of the electron and lattice temperatures after ultrafast optical excitation. Thus, for monochromatic optical excitation, the Fermi smearing of heated conduction electrons induces a nonlinear dependence of the dielectric constant of gold $\epsilon_{\rm Gold} = \epsilon_{\rm m} + \chi^{(3)}_{\rm T}|{\bf e}|^2{\bf e}$, where $\epsilon_{\rm m} = \epsilon^{\prime}_{\rm m} + i\epsilon^{\prime\prime}_{\rm m}$ is the linear dielectric constant of gold, ${\bf E}({\bf r},t) = {\rm Re} [{\bf e}({\bf r}){\rm e}^{-i\omega_0 t}]$ is the electric field of the fundamental TM mode with spatial profile ${\bf e}({\bf r})$ and angular frequency $\omega_0$,
\begin{align}
\chi^{(3)}_{\rm T}(\omega_0)=\frac{1}{2}\epsilon_0\omega_0\epsilon^{\prime\prime}_{\rm m}(\omega_0)\gamma_{\rm T}(\omega_0),
\end{align}
and $\gamma_{\rm T}$ is the thermo-modulational coefficient ($\gamma_{\rm T} \simeq 1.44+i\,0.09$ J$^{-1}$cm$^3$fs and $\epsilon_{\rm m} \simeq -88.35 + i\,5.25$ at $\lambda_0 = 2\pi c/\omega_0 = 1.55\ \mu$m) \cite{MariniNewJPhys2013}. For ultrafast optical pulses the nonlinear polarization resulting from the interband thermo-modulation of gold can be expressed in terms of nonlinear susceptibility and temporal response functions as
\begin{align}
{\bf P}_{\rm NL}(t)= \chi^{(3)}_{\rm T}(\omega_0){\bf e}(t)\int_{-\infty}^{+\infty}dt'\, h_{\rm T}(t-t')|{\bf e}(t-t')|^2, \label{NLGoldPolEq}
\end{align}
where $h_{\rm T}(t) = \Theta(t)(\tau_{\rm th}-\tau_{\rm r})^{-1} ( {\rm e}^{-t/\tau_{\rm th}} - {\rm e}^{- t/\tau_{\rm r})}$ is the temporal response function arising from the two-temperature model and $\Theta(t)$ is the Heaviside step function. Eq. (\ref{NLGoldPolEq}) incorporates the ultrafast nonlinear optical properties of gold, thus enabling the accounting of thermo-modulation in the optical propagation in gold-based plasmonic devices \cite{MariniNewJPhys2013}.

\section{SPP propagation}

SPP pulses propagating in the active plasmonic waveguide depicted in Fig. \ref{DS_model}(a) experience the dispersion and nonlinear effects of the structure. In order to describe ultrafast propagation we consider a generic SPP pulse with optical field ${\bf E}({\bf r},t) = {\rm Re} [\psi(z,t){\bf e}({\bf r}_{\bot}){\rm e}^{-i\omega_0 t}]$, where $\psi(z,t)$ is the field envelope and ${\bf e}({\bf r}_{\bot})$ is the spatial profile of the fundamental TM mode at angular frequency $\omega_0$ \cite{COMSOL}. By developing a perturbative theory accounting for dispersion, absorption, and nonlinearity as small perturbations \cite{MariniPRA2011} , one gets a CGLE for $\psi(z,t)$:
\begin{align} \label{gl1}
& \left\{i\partial_z+i v_{\rm g}^{-1}\partial_t-[ i G_2 + (\beta_2/2) ]\partial_t^2 -i(G_0-l) \right\}\psi(z,t) \nonumber \\
& + \gamma_{\rm Au} \psi (z,t)\int\limits_0^{\infty }{d{t}'h_{\rm T}(t')|\psi(z,t-t')|^2} = 0,
\end{align} 
where $v_{\rm g}\simeq 5.72\times 10^7$ m/s is the group velocity, $\beta_2\simeq-0.40$ ps$^2$/m is the GVD parameter. In the unsaturated gain regime, the frequency dependent gain profile is approximated as a Lorentzian \cite{GPA91}, $G(\omega) \approx G_0-G_2(\omega-\omega_0)^2$, where $G_0\simeq 7.2$ dB/cm is the linear gain provided by Al$_2$O$_3$:Er$^{3+}$ at $\lambda_0 = 2\pi c/\omega_0 = 1.55\,\mu$m and $G_2 \simeq 4.1\times 10^3$\,dB\,fs$^2$/cm is the gain dispersion parameter. $l \simeq 0.5$ dB/mm is the linear loss arising from gold absorption, and
\begin{subequations}
\label{NLCoeffEqs}
\begin{align}
\gamma_{\rm Si}=\frac{\omega_0 \chi^{(3)}_{\rm Si}}{4\epsilon_0 c^2}\frac{\int_{\rm Si ~ area}[2|{\bf e}|^4 +|{\bf e}^2|^2]}{(\int_{\rm Full ~ area} {\rm Re} [{\bf e} \times {\bf h}^*]\cdot\hat{\bf z})^2}, \\
\gamma_{\rm Au}=\frac{\omega_0 \chi^{(3)}_{\rm T}}{\epsilon_0 c^2}\frac{\int_{\rm Au ~ area}|{\bf e}|^4}{(\int_{\rm Full ~ area} {\rm Re} [{\bf e} \times {\bf h}^*]\cdot\hat{\bf z})^2}.
\end{align}
\end{subequations} 
Note that the upper integrals are extended in areas where gold/silicon are located, respectively, $\hat{\bf z}$ is the longitudinal unit vector, ${\bf h} = -i(c/\omega)\nabla\times{\bf e}$ is the rescaled magnetic field of the mode, and $\chi^{(3)}_{\rm Si}$ is the third-order complex nonlinear susceptibility of silicon. We emphasize that identical expressions for the nonlinear coefficients $\gamma_{\rm Si},\gamma_{\rm Au}$ are obtained through the reciprocity theorem \cite{Afshar}. Note that in Eq. (\ref{gl1}) we have disregarded the silicon nonlinearity since, as we have discussed in Sec. 2, $|\gamma_{\rm Si}|<<|\gamma_{\rm Au}|$ in the frequency domain considered in our calculations. The CGLE can be simplified in the moving reference frame leading to the normalized extended CGLE~\cite{Akhmediev}
\begin{align} \label{gl2}
& i\frac{\partial u}{\partial \xi }-\frac{s}{2}\frac{\partial^2u}{\partial\tau^2}-i\left( g_0+g_2\frac{\partial^2}{\partial\tau^2} \right)u +  \\ 
& + i\alpha u + (1+iK)\left(|u|^2 -\tau_{\rm T} \frac{\partial|u|^2}{\partial\tau}\right)u =0 ,\nonumber
\end{align}
where in the convolution in Eq.\,(\ref{gl1}) we have approximated $|\psi(z,t-t')|^2 \simeq |\psi(z,t)|^2 -t'\frac{\partial}{\partial t}|\psi(z,t)|^2$ obtaining $\int\limits_0^{\infty }{dt'h_{\rm T}(t')|\psi(z,t-t')|^2}\simeq |\psi(z,t)|^2 - \tau_{\rm T} \frac{\partial}{\partial t}|\psi(z,t)|^2$ where $\tau_{\rm T}=\tau_{\rm th}+\tau_{\rm r}$, since $\int_0^{+\infty} h_{\rm T}(t')dt' = 1$ and 
$\int_0^{+\infty} t' h_{\rm T}(t')dt'=\tau_{\rm T}$. Such an approximation holds as long as the pulse duration $t_0$ is much longer than the thermal time $\tau_{\rm T}$. The time ($t$) and propagation distance ($z$) variables are normalized as $\tau=(t-zv_{\rm g}^{-1})/t_0$ and $\xi=z/L_{\rm D}$, where $L_{\rm D} =t_0^2/|\beta_2 (\omega_0)|$ is the dispersion length and $s={\rm sgn} (\beta_2)$. The field amplitude ($\psi$) is rescaled as $\psi = u\sqrt{P_0}$, where $u(\xi,\tau)$ is the normalized pulse envelope and $P_0=|\beta_2 (\omega_0 )|/(t_0^2 \,\rm{Re}[\gamma_{\rm Au}])$ is the peak power. Other normalized parameters in Eq. (\ref{gl2}) are the dimensionless gain $g_0=G_0 L_{\rm D}$ and its dispersion $g_2=G_2L_{\rm D}/t_0^2$, linear absorption $\alpha=l L_{\rm D}$ and nonlinear loss $K=\rm{Im}[\gamma_{\rm Au}]/\rm{Re}[\gamma_{\rm Au}]$. 

\begin{figure}[t]
\centering
\begin{center}
\includegraphics[width=0.5\textwidth]{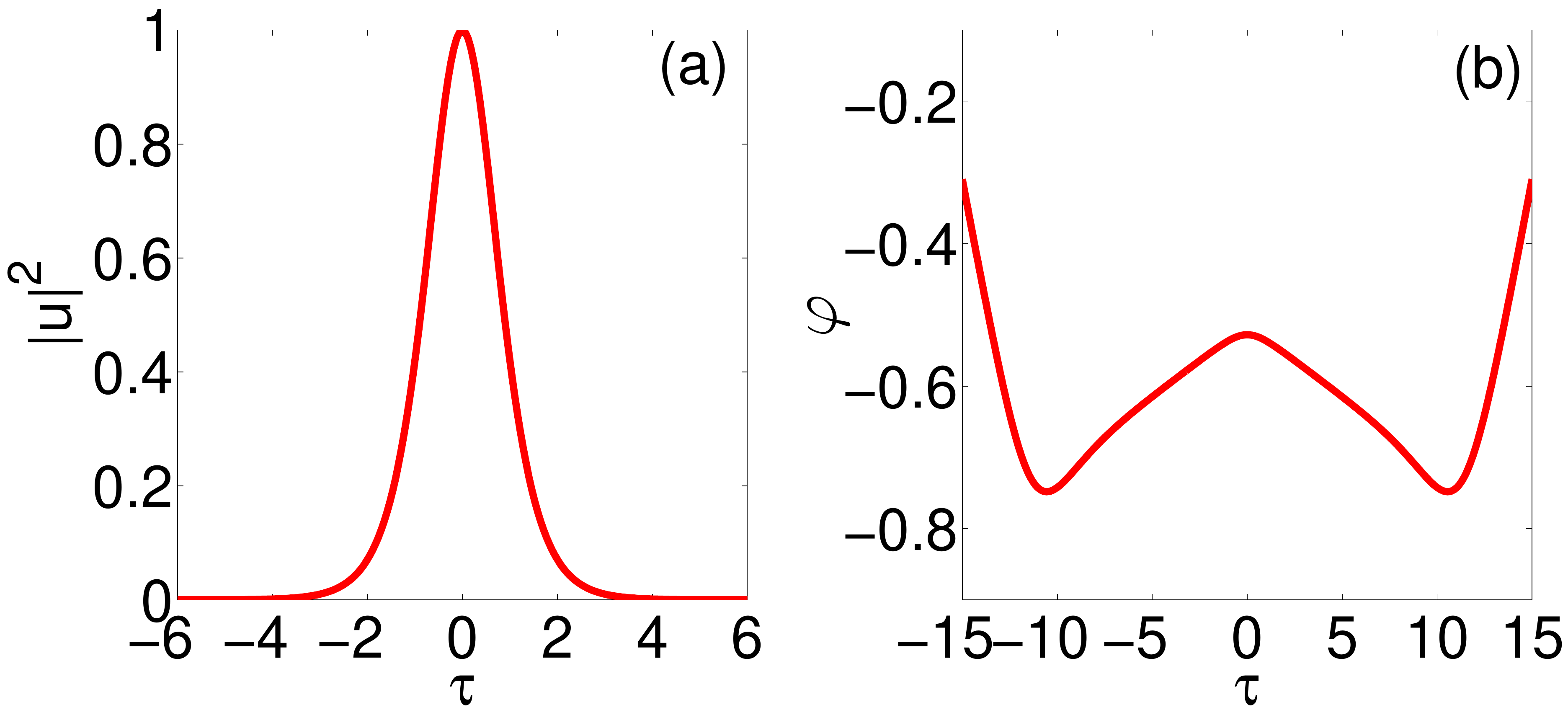}
\caption{(Color online) (a) Intensity $|u|^2$ and (b) phase pattern $\varphi = {\rm atan}\{{\rm Im}[u]/{\rm Re}[u]\}$ of the DPS for $K=0.01,~g_0=0.1,~g_2=0.01$ and $\alpha=0.09$.} 
\label{Fig2}
\end{center}
\end{figure}

\section{Dissipative plasmon solitons}

\begin{figure}[t]
\centering
\begin{center}
\includegraphics[width=0.47\textwidth]{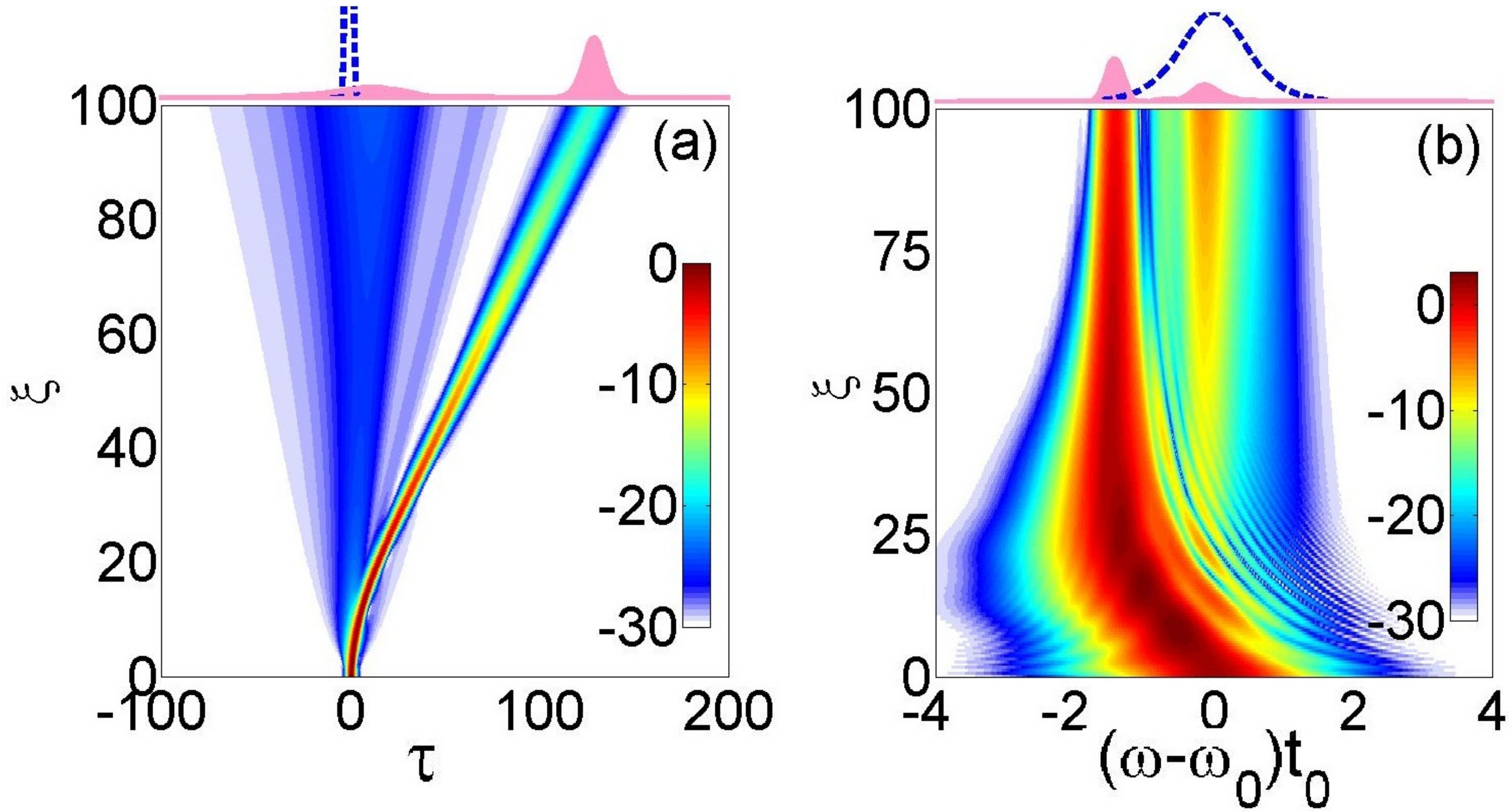}
\caption{(Color online) Evolution of a DPS in the (a) time and (b) frequency domains for $\tau_T=0.2$. The input (dotted trace) and output pulse shapes are also shown in the top panel. The parameters used in the simulations are: $K=0.01,~\widetilde{g_0}=0.01$ and $g_2=0.01$.}  
\label{Fig3}
\end{center}
\end{figure}

For long pulses, neglecting the effect of thermo-modulational delayed nonlinearity (i.e., by setting $\tau_T=0$), there exist particular localized and self-sustaining pulses propagating in the active plasmonic waveguides without changing their intensity shape $|u|^2$. The envelope of such peculiar nonlinear waves is analogous to the well-known Pereira-Stenflo DPSs ~\cite{PS}, explicitly 
\begin{equation} \label{ps1}
u(\xi, \tau)=u_0[{\rm sech}(\eta\tau)]^{(1+ia)}{\rm e}^{i\Gamma\xi},
\end{equation}
where the parameters $u_0, \eta$, $\Gamma$ and $a$ are not independent on each other, but are set by the so-called DPS existence law 
\begin{align} \label{parameters}
    |u_0|^2 = \frac{\widetilde{g_0}}{K}\left[ 1-\frac{sa/2+g_2}{g_2(a^2-1) -sa} \right], 
\end{align}
where $\Gamma=(\eta^2/2)[s(a^2-1)+4ag_2]$, $\eta^2 = \widetilde{g_0}/[g_2(a^2-1) -sa]$, $a=[H-\sqrt{H^2+2\delta^2}]/\delta$, $H=-3[s/2+g_2 K]$, $\delta=[ sK-2g_2]$ and $\widetilde{g_0}=(g_0-\alpha)$. The intensity $|u|^2$ and phase pattern $\varphi = {\rm atan}\{{\rm Im}[u]/{\rm Re}[u]\}$ profiles of the Pereira-Stenflo DPS given by Eq. (\ref{ps1}) are depicted in Fig. \ref{Fig2} for $K=0.01$, $\widetilde{g_0}=0.01$, $g_2=0.01$. Note that the temporal phase gradient is responsible for intrapulse energy flow enabling unperturbed intensity propagation in a temporally inhomogeneous amplification mechanism. While such DPSs in principle should propagate without pulse distortion, owing to the thermo-modulational nonlinear perturbation they decelerate in time and red-shift accordingly [see Figs.~\ref{Fig3} (a,b) where we plot the temporal and spectral evolutions of the perturbed DPS in the presence of thermo-modulational nonlinearity for the same parameters of Fig. \ref{Fig2} and $\tau_{\rm T}=0.2$]. Indeed, as a consequence of delayed mechanism of the thermo-modulational nonlinear response, blue components of the spectrum are quenched, leading to a self-induced soliton spectral red-shift that slows down the DPS considerably. We further observe that the red-shift saturates after a particular propagation distance threshold around $\xi=30$ owing to the finite bandwidth of the gain medium accounted by the linear gain dispersion parameter $g_2$. The numerical results presented in Fig. \ref{Fig3} are obtained by a split-step Fourier method embedding a fourth-order Runge-Kutta algorithm \cite{GPA} enabling us to solve numerically Eq. (\ref{gl2}) providing as input pulse the DPS given by Eqs. (\ref{ps1},\ref{parameters}).

\section{Variational analysis}

As we have discussed above, the unaltered DSP nonlinear propagation holds only for small distances where thermo-modulational nonlinear dynamics [accounted by the term $\tau_{\rm T}\frac{\partial|u|^2}{\partial\tau}$ in in Eq. (\ref{gl2})] does not come into play. In order to gain physical insight in the complex nonlinear dynamics illustrated in Fig. \ref{Fig3}, here we investigate the effect of thermo-modulational nonlinearity by developing a soliton perturbation theory \cite{GPA} and considering the delayed mechanism ($\tau_{\rm T} \frac{\partial|u|^2}{\partial\tau}$) as a small perturbation to the main cubic nonlinear term ($|u|^2$). Assuming anomalous dispersion $(s=-1)$ we recast the CGLE as a perturbed nonlinear Schr\"{o}dinger equation
\begin{equation} \label{nls2}
i\frac{\partial u}{\partial\xi}+\frac{1}{2}\frac{\partial^2u}{\partial\tau^2}+|u|^2u = i\epsilon(u),
\end{equation}
where 
\begin{equation} \label{eps1}
\epsilon(u)=\left(\widetilde{g_0}+g_2\frac{\partial^2}{\partial\tau^2}-K|u|^2-i\tau_{\rm T} (1+iK) \frac{\partial|u|^2}{\partial\tau}\right)u
\end{equation}
is considered as a small perturbation ($|\epsilon(u)|<<|u|^3$) depending on $u$, $u^*$ and their temporal derivatives. Thus, in order to develop the soliton perturbation theory, we assume that the effect of the perturbation $\epsilon(u)$ is to modify adiabatically the DPS $u(\xi,\tau) = u_0{\rm sech}^{1+ia}[\eta(\tau-\tau_{\rm p})] {\rm e}^{i\phi-i\Omega(\tau-\tau_{\rm p})}$, where the parameters $u_0,~\eta,~\tau_p,~\phi,~a$ and $\Omega$ become smooth functions of $\xi$. Introducing the Lagrangian density ${\cal L}_{\rm D} = (i/2)(u u_{\xi}^* -u^* u_\xi )-(1/2)(|u|^4-|u_\tau|^2)+i(\epsilon u^*-\epsilon^* u)$ and integrating it over $\tau$, one gets the total Lagrangian as $L=\int_{-\infty}^\infty{\cal L}_D d\tau$, explicitly given by 
\begin{align}
& L = \frac{2u_0^2}{\eta}(\frac{\partial\phi}{\partial\xi} + \Omega \frac{\partial\tau_{\rm p}}{\partial\xi}) -\frac{au_0^2}{\eta^2}\frac{\partial\eta}{\partial\xi} + C \frac{u_0^2}{\eta}\frac{\partial a}{\partial \xi } +  \\
& + \frac{\eta u_0^2}{3}(1+a^2) + \frac{u_0^2}{\eta}\left(\Omega^2-\frac{2}{3}u_0^2\right) + 2{\rm Re} \int_{-\infty }^{\infty}[i\epsilon u^*]\,d\tau,\nonumber 
\end{align}
where $C= \ln 4 - 2$. Such Lagrangian leads to the reduced Euler--Lagrange equations of motion for the DPS parameters accounting for the evolution of the pulse energy 
$E=\int_{-\infty }^{+\infty}{|u|^2 d\tau}$, temporal position $\tau_{\rm p}$, frequency shift $\Omega$, inverse duration $\eta$, chirp $a$ and phase $\phi$ through
\begin{subequations}\label{SolPerThEqs}
\begin{align}
& \frac{dE}{d\xi } =\frac{2}{3}( 3\widetilde{g_0}-K\eta E)E- \frac{2}{3}g_2[(1+a^2)\eta^2+3\Omega^2]E , \label{var7}\\ 
& \frac{d\tau_{\rm p}}{d\xi} =-(1+2g_2a)\Omega -\frac{1}{3}\tau_{\rm T} K E \eta,  \label{var8}\\
& \frac{d\Omega }{d\xi } =-\frac{4}{3}g_2(1+a^2)\Omega\eta^2-\frac{4}{15}\tau_{\rm T}(1+aK)E\eta^3, \label{var9}\\ 
& \frac{d\eta }{d\xi } =\frac{2}{9} (3a\eta-EK)\eta^2-\frac{4}{9}(2-a^2)g_2\eta^3 ,\label{var10}\\
& \frac{da}{d\xi } =\frac{1}{3}(1+aK)E\eta -\frac{2}{3} (1+ag_2)(1+a^2)\eta^2.\label{var11} 
\end{align}
\end{subequations}
These novel equations enable to evaluate the evolution of the DPS under thermo-modulational perturbation, which is accounted by the terms containing the effective thermal time $\tau_{\rm T}$. From the equations above we observe that thermo-modulational nonlinearity affects mainly the temporal position $\tau_{\rm p}$ and the frequency shift $\Omega$ [see Eqs.(\ref{var8},\ref{var9})]. In particular, Eq.(\ref{var9}) indicates that the perturbation leads to a strong spectral red-shift of the DPS. We also observe that, while most of soliton parameters are dynamically coupled, the evolution of the global phase $\phi$ remains uncoupled from other soliton parameter, and thus we neglect it in the following discussion.

\begin{figure}[t]
\centering
\begin{center}
\includegraphics[width=0.5\textwidth]{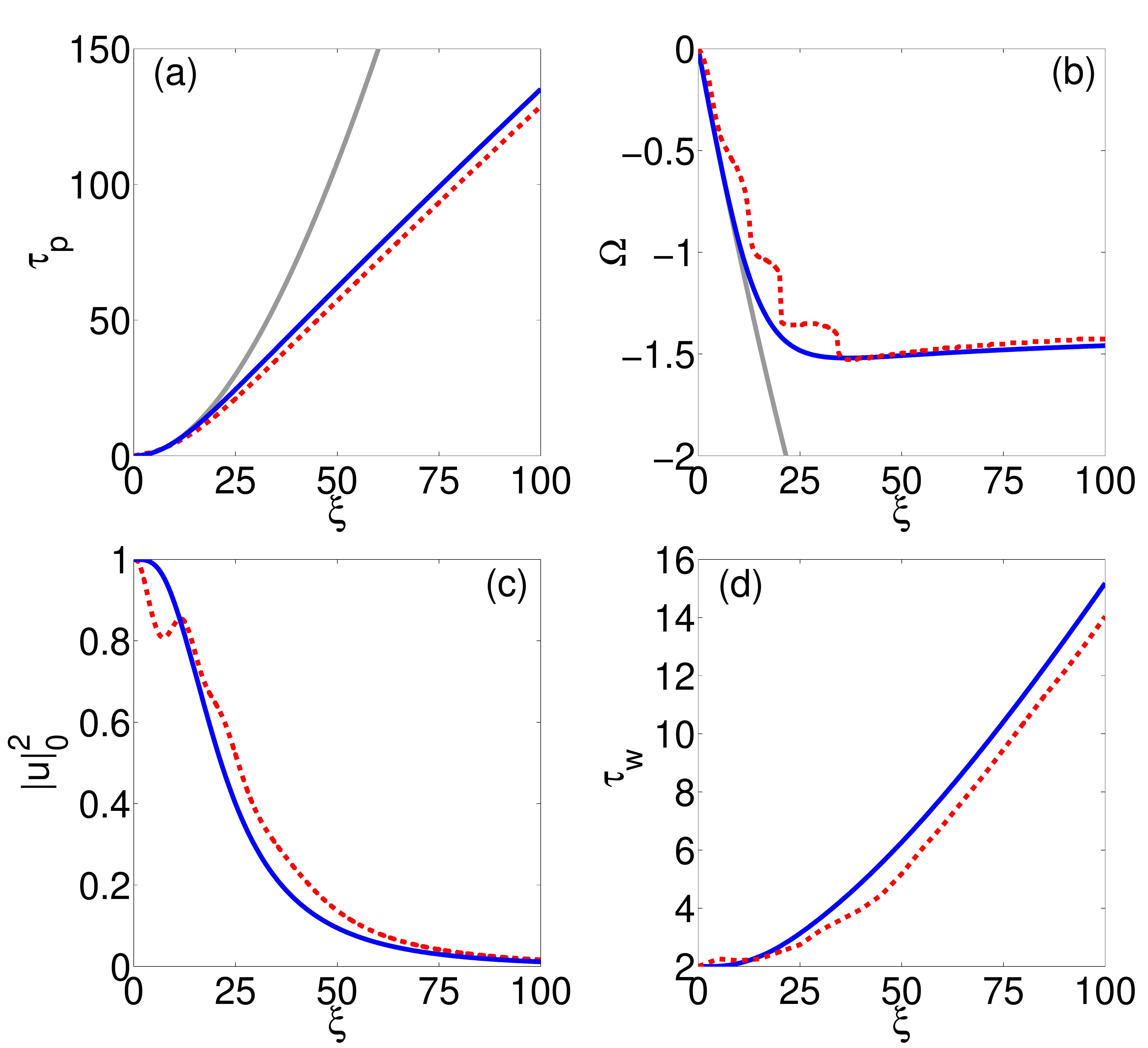}
\caption{(Color online) (a) Dependence of the temporal position $\tau_{\rm p}$ and (b) frequency redshift $\Omega$ over the propagation distance. Such temporal and frequency shifts are accompanied in the dynamics by variations of the peak intensity $|u_0|^2$ and pulse width $\tau_{\rm w}=2/\eta$, which are illustrated in (c) and (d), respectively. Solid blue lines indicate results from the variational theory while dashed red lines correspond to numerical data from the full simulation of Eq. (\ref{gl2}). The gray lines in (a) and (b) indicate analytical predictions provided by Eq.\,(\ref{nls5}) and Eq.\,(\ref{nls4}) respectively. The parameters used in the simulations are $K=0.01$, $\widetilde{g_0}=0.01$, $g_2=0.01$, and  $\tau_{\rm T}=0.2$. } 
\label{Fig4}
\end{center}
\end{figure}

\section{Discussion}

As we have discussed in the previous section, in order to describe the evolution of soliton parameters through the variational approach, in principle one needs to solve Eqs. (\ref{SolPerThEqs}) numerically. In order to gain physical insight, we here assume that the DPS profile changes slightly [which we observe to hold for propagation distances $\xi<20$, see Fig. \ref{Fig2}] 
and approximate $\eta$ and $a$ to the initial DPS constants. This assumption enables us to integrate Eq. (\ref{var9}) analytically and reduce the evolution of the frequency redshift $\Omega$ to a close form
\begin{equation} \label{nls4}
\Omega(\xi) \simeq -\Omega_{\rm T} (1-{\rm e}^{-\rho \xi}),
\end{equation}
where $\Omega_{\rm T}=2\tau_{\rm T} u_0^2(1+aK)/[5g_2(1+a^2)]$ and $\rho=4g_2(1+a^2)\eta^2/3$. This simple analytical expression describes how the DPS redshift increases with propagation distance $\xi$ initially and saturates for long distances to the final value $\Omega(\xi)\rightarrow-\Omega_R$ for $\xi\gg 1$. Under the same assumptions, we integrate also Eq. (\ref{var8}) obtaining an analytical epxression for the temporal shift
\begin{equation} \label{nls5}
\tau_{\rm p}(\xi) \simeq \Omega_{T}(1+2g_2a)[(1-\mathcal{P}_{\rm T})\xi-\rho^{-1}(1-{\rm e}^{-\rho\xi})],
\end{equation}
where $\mathcal{P}_{\rm T} = (5/3)g_2 K(1+a^2)/[(1+2g_2 a)(1+aK)]$. Thus, our analytical predictions indicate that while redshift saturates for $\rho\xi\gg 1$, $\tau_p$ does not saturate and keeps increasing linearly with $\xi$ for $\rho\xi\gg 1$. In Fig. \ref{Fig4} we plot the evolution of temporal position $\tau_{\rm p}$, spectral shift $\Omega$, peak intensity $|u|_0^2$ and pulse width $\tau_{\rm w} =2/\eta$ of a perturbed DPS with parameters indicated in the caption of Fig. \ref{Fig4}. Dashed lines illustrate results of full numerical simulations of the perturbed CGLE provided by Eq. (\ref{gl2}),  full blue lines represent results of soliton perturbation theory provided by the numerical solution of Eqs. (\ref{SolPerThEqs}), while grey lines indicate analytical predictions. Note that all features are well-predicted by our variational method for the propagation distances considered. We emphasize that such results are rigorous for $\tau_{\rm T}<<1$ and lead to good qualitative predictions when $\tau_{\rm T}$ remains relatively small. In conditions where $\tau_{\rm T}>>1$, which hold for femtosecond pulses, such a perturbative picture is inapplicable and the DPS does not evolve adiabatically but rather breaks immediately. Finally, we emphasize that the mismatch between numerical variational predictions and analytical approximations arises from the assumptions that $\eta$, $a$ and $E$ are unaffected by thermo-modulational nonlinearity, while we observe that $E$ gets reduced by more than $50\%$. Thus, even though our closed form expressions provide much physical insight and describe well pulse dynamics qualitatively for short distances, the inclusion of the dynamical evolution of $a$, $\eta$ and $E$ (or $u_0^2$) is essential for accurate results.

\section{CONCLUSIONS}

In conclusion, we have developed a theoretical model enabling the description of the spatio-temporal dynamics of light emitted in an active plasmonic waveguide composed of a thin film of gold sandwiched by two silicon layers immersed in externally pumped Al$_2$O$_3$:Er$^{3+}$. Our model consists of a generalized cubic Ginzburg-Landau equation for the optical field envelope that is able to account for the thermal modulation of gold optical properties. We observe that heating leads to the existence of peculiar dissipative plasmon solitons that redshift and decelerate while propagating inside the active waveguide. Since our formalism is applicable for a variety of distinct active plasmonic waveguides, we envisage that our results are relevant for the design of plasmonic amplifiers and pulse shaping at the nanoscale.

\section*{Acknowledgements}

A.S. acknowledges MHRD, India for a research fellowship. A.M. acknowledges support from the Rita Levi Montalcini Fellowship (grant number PGR15PCCQ5) funded by the Italian Ministry of Education, Universities and Research (MIUR).

\end{document}